\documentclass{article}

\usepackage{epstopdf}

\usepackage{longtable}

\begin{document}

\setlength{\baselineskip}{5mm}

\noindent{\large\bf
The extended Heine-Stieltjes polynomials associated with a special LMG model

}\vspace{4mm}

\noindent{
Feng Pan,$^{a,b}$ Lina Bao,$^a$ Liyuan Zhai,$^a$ Xiaoyue Cui,$^{a}$ and J. P. Draayer$^b$
}\vspace{1mm}

\noindent{\small
{\small$^a$Department of Physics, Liaoning Normal University, Dalian
116029, P. R. China,
\\$^b$Department of Physics and Astronomy,
Louisiana State University, Baton Rouge, LA 70803-4001, USA.}
}\vspace{4mm}

\noindent {\bf Abstract}:~{\small The extended Heine-Stieltjes polynomials  associated with a
special Lipkin-Meshkov-Glick (LMG)
model corresponding to the standard two-site Bose-Hubbard model
are derived based on the Stieltjes correspondence. It is shown that
there is a one-to-one correspondence between zeros of this new polynomial
and solutions of the Bethe ansatz equations for the LMG model.
A one-dimensional classical electrostatic analogue corresponding
to the special LMG model is established according to Stieltjes early work.
It shows that any possible configuration of equilibrium positions
of the charges in the electrostatic problem corresponds
uniquely to one set of roots of the Bethe ansatz equations for the LMG model,
and the number of possible configurations of equilibrium positions of the charges
equals exactly to the number of energy levels in the LMG model.
Some relations of sums of powers and inverse powers of zeros of the new polynomials
related to the eigenenergies of the LMG model are derived.}


\bigskip
\section{Introduction}

It is well-known from a number of studies [1-3]
that the Lipkin-Meshkov-Glick (LMG) model [4-6] has a rich phase structure that
is related to a broad range of phenomena in a variety of physical applications,
such as analyses of spin
systems [7], Bose-Einstein condensates [8], etc.
Furthermore, it has been proven [9,10] that the model is exactly solvable.
A special case of the theory is related
to the standard two-site Bose-Hubbard model [11] with a Hamiltonian
that is equivalent to that of a paramagnet of the easy-axis type in
a transverse magnetic field [12]. The quantum dynamics and the energy gap
of the model were studied in [13-14]. The quantum critical behavior
and a special case of the model were discussed in [15-16].
As shown in [9-11, 12-16], exact solutions of the model
can be determined using the Gaudin-Richardson or Bethe ansatz method, in which
eigenstates of the system are written in terms of a product of spectral
parameter-dependent operators. The unknown spectral parameters must satisfy
a set of coupled non-linear equations, called the Bethe ansatz equations (BAEs).
Solutions of the BAEs simultaneously determine the eigenenergies and the
corresponding eigenstates [9-11].

It should be noted that according to
the early work of Stieltjes [17-20] there is an important
one-to-one correspondence between every set of BAEs
and a set of orthogonal polynomials. And furthermore,
as shown by Szeg\"{o} [21], the Stieltjes correspondence
can be used to define a large class of classical orthogonal
polynomials. Roots of these BAEs are zeros of
the corresponding polynomials, which can be interpreted
as stable equilibrium positions for a set of free charges
in an external electrostatic field. This link between Richardson's BCS pairing
model for nuclei and the corresponding electrostatic problem was
established in [22] based on an earlier unpublished preprint of Gaudin,
which was then made clearer in [23]. A much more general approach
to the pairing model was shown in [24] and [25].
The purpose of this paper is
to show that the Stieltjes correspondence related to the solutions of
a special LMG model corresponding to the standard two-site Bose-Hubbard model
gives rise to a new polynomials, which is now called  the
extended Heine-Stieltjes polynomials .

\section{Stieltjes correspondence}

There is a large class of polynomials, called Heine-Stieltjes polynomials $y(z)$,
which satisfy a second-order Fuchsian equation

$$A(x)y^{\prime\prime}(x)+B(x)y^{\prime}(x)+V(x)y(x)=0,\eqno(1)$$\\
where $A(x)$  is a polynomial of degree $m$ with $A(x)=\prod_{i=1}^{m}(x-a_{i})$,
$B(x)$ is a polynomial of degree $m-1$ such that for a set of real
positive parameters $\{\gamma_{i}\}$ and another set of real
parameters $\{a_{i}\}$

$$B(x)/A(x)=\sum^{m}_{i=1}{\gamma_{i}\over{x-a_{i}}},\eqno(2)$$\\
and $V(x)$ is an unknown, but to be determined polynomial of degree
$m-2$ that is allowed to depend on the solution $y(x)$.
The latter are often called Van Vleck polynomials [26].
The case with $m=2$ corresponds to the hypergeometric differential equation,
while the case with $m=3$ corresponds to the Heun equation [27-28].

Such Heine-Sieltjes polynomials and properties of their zeros have been studied
extensively. If the polynomials $A(x)$ and $B(x)$
are algebraically independent, i.e., they do not satisfy any algebraic
equation with integer coefficients, Heine proved that
for every integer $k$ there exist at most
$\sigma(k)=(k+m-2)!/((m-2)! k!)$ different  Van Vleck polynomials $V(x)$
such that $y(x)$ has a polynomial solution of degree $k$.
As summarized in
Szeg\"{o}'s work on orthogonal polynomials [21], for every set of nonnegative integers
$(k_1, k_2,\cdots, k_m)$, there are uniquely determined
real values of the parameters for the Van Vleck polynomials $V(x)$ such that $y(x)$
has a polynomial solution
with $k_j$ simple zeros in the open interval $(a_{j-1}, a_j)$
for each $j=1,2,\cdots, m$. The polynomial
$y(x)$ is uniquely determined  up to a constant factor, and has the degree
$k=k_1+\cdots k_m$.  As noted by Volkmer [29], this existence and uniqueness
theorem can also be derived from the general Klein oscillation theorem for multi-parameter
eigenvalue problems shown in [30].

If $y(x)$ is a polynomial of degree $k$ with simple zeros
$\{x_{1},x_{2},\cdots,x_{k}\}$, one may write $y(x)$ as

$$y(x)=\prod_{j=1}^{k}(x-x_{j}).\eqno(3)$$\\
Thus, one can easily check that $y(x)$
will satisfy

$${y^{\prime\prime}(x_{i})\over{y^{\prime}(x_{i})}}
=\sum_{j\neq i}{2\over{x_{i}-x_{j}}}\eqno(4)$$\\
at any zero $x_{i}$. Combining (1), (2), and (4),
one obtains the following important relations among the zeros:

$$\sum_{j\neq i}^{k}{2\over{x_{i}-x_{j}}}+\sum^{m}_{\mu=1}{\gamma_{\mu}\over{x_{i}-a_{\mu}}}=0\eqno(5)$$\\
for $i=1,2,\cdots, k$. It should be noted that the BAEs frequently appearing in search for
exact solutions of quantum many-body problems, such as those associated with Gaudin type
systems, are similar to the relations shown in (5). The link between Richardson's BCS
pairing model for nuclei and the corresponding electrostatic problem was
investigated in [22] based on an earlier unpublished preprint of Gaudin,
which was then made clearer in [23]. A much more general approach
to the pairing model was shown in [24] and [25].
Roots of the BAEs (5) simultaneously
determine the eigenenergies and eigenstates of the corresponding quantum many-body problem.
As an alternative, roots of the BAEs may also be calculated as zeros of the
corresponding polynomial $y(x)$. In this way, a link between
solutions of the Gaudin type for quantum many-body problems and
the corresponding polynomials is established.

\section{Special LMG model and Bethe ansatz solutions}

As a simple extension of the Stieltjes correspondence, we revisit the Bethe ansatz
solutions for a special LMG model corresponding to the standard two-site Bose-Hubbard model
studied in [11, 13-15], for which the Hamiltonian is

$$\hat{H}=-t(c^{\dagger}d+d^{\dagger}c)+U(c^{\dagger}cc^{\dagger}c+
d^{\dagger}dd^{\dagger}d),\eqno(6)$$\\
where $c^{\dagger}$ ($c$) and $d^{\dagger}$ ($d$) are boson creation
(annihilation) operators, the parameters $t$ and $U$ in the Bose-Hubbard model
are related to the Josephson coupling and the charging energy, respectively.
Following [11, 15], we use the unitary transformation
for the boson operators with

$$c = \sqrt{1\over{2}}(a -\dot{\imath}b),~~~ d =\sqrt{1\over{2}}(a +{\dot{\imath}}b).\eqno(7)$$\\
Then, the Hamiltonian (6) can be rewritten in terms of $a$- and $b$-boson operators
as

$$\hat{H}=t(b^{\dagger}b-a^{\dagger}a)-{1\over{2}}U
S^{+}(0)S^{-}(0)+U\hat{n}^2,\eqno(8)$$\\
where $\hat{n}=a^{\dagger}a+b^{\dagger}b$ is the operator for the total number
of bosons in the system, and

$$S^{+}(0) = b^{\dagger2} + a^{\dagger2},
~~S^{-}(0) = b^{2} + a^{2}\eqno(9)$$\\
are boson pairing operators.

Let $\vert \nu_1,\nu_2\rangle=a^{\dagger\nu_1}b^{\dagger\nu_2}\vert 0\rangle$
with $\nu_{i}=0$ or $1$ for $i=1, 2$
be the $a$- and $b$-boson pairing
vacuum state satisfying

$$a^{2}\vert \nu_1,\nu_2\rangle=b^{2}\vert \nu_1,\nu_2\rangle=0.\eqno(10)$$\\
To diagonalize the Hamiltonian, we use the algebraic Bethe ansatz
which implies that eigenvectors of (8) may be expressed as

$$\vert n,~\zeta,\nu_{1},\nu_{2}\rangle=
S^{+}({x}^{(\zeta)}_{1})S^{+}({x}^{(\zeta)}_{2})\cdots
S^{+}({x}^{(\zeta)}_{k})\vert \nu_1,\nu_2\rangle\eqno(11)$$\\
with $n=2k+\nu_{1}+\nu_{2}$, and

$$S^{+}({x}^{(\zeta)}_{i})={a^{\dagger 2}\over{x_{i}^{(\zeta)}+1}}+{b^{\dagger
2}\over{
x^{(\zeta)}_{i}-1}},\eqno(12)$$\\
in which $x^{(\zeta)}_{i}$ ($i=1,2,\cdots,k$)
are spectral parameters to be determined,
and $\zeta$ is an additional quantum number for distinguishing
different eigenvectors with the same quantum number $k$.
It can then be verified by using the corresponding
eigen-equation that (11) is a solution when the
spectral parameters $x_{i}^{(\zeta)}$ ($i=1,2,\cdots, k$)
satisfy the following set of BAEs:

$$U\left( {2\nu_{1}+1\over{x_{i}^{(\zeta)}+1}}+
  {2\nu_{2}+1\over{x_{i}^{(\zeta)}-1}}\right)+2t
  +4U\sum_{j(\neq i)}{1
  \over{x_{i}^{(\zeta)}-x_{j}^{(\zeta)}}}=0\eqno(13)$$\\
for $i=1,2,\cdots,k$, with the corresponding eigen-energy given by

$$E^{(\zeta)}_{n}=2t\sum_{i=1}^{k}{{x_{i}^{(\zeta)}}}+
t(\nu_2-\nu_1)+Un^{2}.\eqno(14)$$\\
Hence, once the $\zeta$-th roots $\{x_{i}^{(\zeta)}\}$ are obtained
from Eq. (13), the  eigenenergy and the corresponding
eigenstate are thus determined according to (14), (11), and (12).

\section{The extended Heine-Stieltjes polynomials
associated with solutions of the LMG model}

In order to compare to the Stieltjes correspondence,
we assume $t>0$ and $U\neq 0$, and set

$$\alpha=\nu_{1}+1/2,~~~\beta=\nu_{2}+1/2,~~~ \gamma=t/U.\eqno(15)$$\\
Then, the BAEs (13) becomes

$$\left( {\alpha\over{x_{i}^{(\zeta)}-1}}+
  {\beta\over{x_{i}^{(\zeta)}+1}}\right)+\gamma
  +\sum_{j(\neq i)}{2
  \over{x_{i}^{(\zeta)}-x_{j}^{(\zeta)}}}=0.\eqno(16)$$\\
When $U>0$, all parameters $\alpha$, $\beta$ and $\gamma$
are always positive. When $U<0$, we interchange the boson
operator $a^{\dagger}$ with $b^{\dagger}$ in (12). Then, one can easily
verify that such a change is equivalent to interchange $\nu_{1}$ with
$\nu_{2}$ and $t\rightarrow-t$ in the BAEs (13), which leads to
the final BAEs for the $U<0$ case that is the same as (16)
with $\alpha\rightleftharpoons\beta$ and keeping
$\alpha>0$, $\beta>0$, and $\gamma>0$. Hence, it is sufficient
to consider the case for $U>0$ only with BAEs given by (16).

Although the $\gamma=0$ result is trivial for the LMG model,
according to the Stieltjes correspondence, the polynomial
corresponding to (16) in this case is the Jacobi polynomial
$P^{(\alpha-1,\beta-1)}_{k}(x)$  satisfying the well-known
differential equation

$${d^{2}\over{dx^{2}}}P^{(\alpha-1,\beta-1)}_{k}(x)+\left( {\alpha\over{x-1}}+
  {\beta\over{x+1}}\right){d\over{dx}}P^{(\alpha-1,\beta-1)}_{k}(x)
  -{k(k+\alpha+\beta-1)\over{x^{2}-1}}P^{(\alpha-1,\beta-1)}_{k}(x)=0.\eqno(17)$$\\
In this case, the Van Vleck polynomial $V(x)$ is trivially a $k$-dependent
constant. Hence, there is only one set of zeros of $P^{(\alpha-1,\beta-1)}_{k}(x_{i})$
with $i=1,2,\cdots, k$, satisfying the Bethe ansatz equation (16).

The case with $\gamma\neq0$ is non-trivial.
According to (1)-(5), we write a differential
equation corresponding to the Bethe ansatz equation (16) as

$${d^{2}y(x)\over{dx^{2}}}+\left( {\alpha\over{x-1}}+
  {\beta\over{x+1}}+\gamma\right){dy(x)\over{dx}}
 +{V(x)\over{x^{2}-1}}y(x)=0\eqno(18)$$\\
with the corresponding polynomials $A(x)=x^2-1$ and
$B(x)=\gamma x^2+(\alpha+\beta)x+\alpha-\beta-\gamma$
shown in (1). In contrast to the Heine-Sieltjes equation (1),
however, in this case the polynomial $B(x)$ is of the same degree
as that of $A(x)$. Therefore, the polynomials
$y(x)$ determined by (18) should be similar to but different
from those of Heine-Sieltjes type.

In search for polynomial solutions of (18), we write
$$y_{k}(x)=\sum_{j=0}^{k}b_{j}x^{j}.\eqno(19)$$
Substitution of (19) into (18) yields
the condition to determine the corresponding
polynomial $V(x)$ with

$$V(x)=-\gamma kx+f,\eqno(20)$$
where $f$ is an undetermined constant depending on $k$,
together with the expansion coefficients $b_{j}$, satisfying
the following four-term relations:

$$\left( j(\alpha+\beta+j-1)+f\right)b_{j}=
(j+2)(j+1)b_{j+2}+(j+1)(\beta-\alpha+\gamma)b_{j+1}+
\gamma(k-j+1)b_{j-1}\eqno(21)$$\\
with $b_{j}=0$ for $j\leq-1$ or $j\geq k+1$, which is
equivalent to the following eigen-equation for $f$
with

$${\bf F}{\bf b}=f{\bf b},\eqno(22)$$
where

$${\bf F}=\left(
\begin{tabular}{c}
~~0~~~~~~$(\beta-\alpha+\gamma)$~~~~~~~~~~~~2 ~~~~~~~~$0\cdots$~~~~~~~
~~~~~~~~~~~~~~~~~~~~~~~~~~~~~~~~~~\\\\
$k\gamma$~~~~$-(\alpha+\beta)$~~~~~~$2(\beta-\alpha+\gamma)$~~~~~~~~$6$~~~~~~~~~$0\cdots$~~~~
~~~~~~~~~~~~~~~~~~~~~~\\\\
~~~~0 ~~~~~~$(k-1)\gamma$~~ ~$-2(\alpha+\beta+1)$ ~~$3(\beta-\alpha+\gamma)$~~$12$~~~$0\cdots$
~~~~~~~~~~~~~~~~~~~~\\\\
~~~~~~~~~~~~~~~$\ddots$~~~~~~~~~~~~~~~~~~~~$\ddots$~~~~~~~~~~~~$\ddots$~~~~~~~~~~~~$\ddots$
~~~~~~~~~~~~\\
~~~~~~~~~~~~~~~~~~~~~~~~~~~$\ddots$~~~~~~~~~~~~~~~~~~$\ddots$~~~~~~~~~~~~~~~$\ddots$~~~~~~$k(k-1)$\\\\
~~~~~~~~~~~~~~~~~~~~~~~~~~~~~~~~~~~~~~~~~$2\gamma$~~~~$-(k-1)(\alpha+\beta+k-2)$~~~$k(\beta-\alpha+\gamma)$ \\\\
~~~~~~~~~~~~~~~~~~~~~~~~~~~~~~~~~~~~~~~~~~~~~~~~~~~~~~~~$\gamma$ ~~~~~~~~~~~~~~~~~~$-k(\alpha+\beta+k-1)$\\
\end{tabular}\right),\eqno(23)$$\\
and the transpose of  ${\bf b}$ is related to the expansion coefficients $\{b_{j}\}$ with
 ${\bf b}^{T}=\left( b_{0},b_{1},\cdots,b_{k-1},b_{k}\right)$.
It is clear that the matrix ${\bf F}$ should have $k+1$ eigenvalues labeled as
$f^{(\zeta)}$ ($\zeta=1,2,\cdots,k+1$), of which
the corresponding eigenvector ${\bf b}^{(\zeta)}$ determines the
polynomial $y^{(\zeta)}_{k}(x)$ according to (19). The number of solutions for $f$
is indeed the same as that of the eigenstates of the standard two-site
Bose-Hubbard Hamiltonian (8) when $\gamma\neq 0$.
The polynomials (19) with $\gamma\neq0$ determined by (21)
are called the extended Heine-Stieltjes polynomials, which,
in general, can also be obtained from the Riccati differential
equation studied in [31-32] or from relevant Bethe ansatz equations
given in [33] though these authors did not intend to do so.
It should be noted that the polynomial approach shown above
is similar to that studied in [34-36] for quasi-exactly solvable sextic anharmonic oscillator problem
and ${\cal PT}$-symmetric quantum mechanics. Since the Bethe ansatz equations
of the quasi-exactly solvable sextic anharmonic oscillator problem
and ${\cal PT}$-symmetric quantum mechanics are different from
those for the special LMG model, the resultants should belong to
different types of extended Heine-Stieltjes polynomials.

As is well known, the advantage of the Bethe ansatz method
for Gaudin type systems lies in the fact that the huge matrix
in the Fock subspace is greatly reduced, especially for the Gaudin-Richardson
pairing model [22-25].
However, the non-linear Bethe ansatz  equations similar to (16)
are very difficult to be solved numerically,
especially for large size systems. There are several attempts to overcome
this difficulty. The approach via Riccati differential
equation shown in [31-32] is one of them.
Actually, if the polynomials can be derived recursively
similar to (21) for the LMG model, it should be much easier to
determine zeros from the polynomials than to solve a set
of BAEs with a set of variables because there is only one
variable in the polynomials. In order to make this point clear,
let us take a simple nontrivial example with $k=2$ and
$\alpha=\beta=\gamma=1/2$ corresponding to $\nu_1=\nu_2=0$
and $t/U=1/2$. In this case, the three polynomials $y_2(x)$ given by (19) with $b_{2}=1$
can easily obtained from the four-term recurrence relations (21)
as listed in Table 1.  Since we set $b_{k}=1$ in $y(x)$,
the coefficient $b_{k-1}$ must equal to
negative sum of zeros of $y(x)$ with
$b_{k-1}=-\sum_{i=1}^{k}x_{i}$.
Therefore, the solution corresponding to the largest $b_{k-1}$
is that for the ground state of the system considered;
that corresponding to the next largest $b_{k-1}$
is that of the first excited state; and so on. One can check
that these are indeed the case as shown in Table 1.

\begin{table}[ht]
\caption{The extended Heine-Stieltjes polynomials $y_2(x)$
for the LMG model with $\alpha=\beta=\gamma=1/2$.}
\label{energy} \vspace*{-6pt}
\begin{center}
\begin{tabular}{cccc}\hline\hline
Zeros of the polynomial &$x_1+x_{2}$ &The Polynomial $y_2(x)$ &$f$ \\ \hline
$x_{1}=-7.2904,~x_{2}=-1.7379$ &$-9.0283$ &$12.6700+9.0283x+x^2$&$~~0.5141$\\
$x_{1}=-5.7052,~x_{2}=~0.5612$ &$-5.1440$ &$-3.2020+5.1440x+x^2$ &$-1.4280$\\
$x_1=-0.6036,~x_{2}=~0.7759$ &$~~0.1723$ &$-0.4684-0.1723x+x^2$&$-4.0861$\\
\hline \hline
\end{tabular}
\end{center}
\end{table}

According to the Stieltjes results, an electrostatic interpretation of the location of
zeros of the new polynomial $y(x)$ may be stated as follows.
Put two positive fixed charges $\alpha/2$ and $\beta/2$ at $-1$ and $+1$
along a real line, respectively,
and allow $k$ positive unit charges to move freely along the real line
under such situation
together with a uniform
electric field with strength $-\gamma/2$. Therefore, up to a constant, the total energy functional
$U(x_{1},x_{2},\cdots, x_{k})$ may be written as

$$U(x_{1},x_{2},\cdots, x_{k})=-{\gamma\over{2}}\sum_{i}x_{i}
-{\alpha\over{2}}\sum_{i}\ln{\vert x_{i}-1\vert}-{\beta\over{2}}\sum_{i}\ln{\vert x_{i}+1\vert}
-\sum_{1\leq i<j\leq k}\ln\vert x_{i}-x_{j}\vert.\eqno(24)$$\\
In this case, there are $k+1$ different configurations for the position
of these $k$ charges $\{x^{(\zeta)}_{1},\cdots, x^{(\zeta)}_k\}$
with $\zeta=1,2,\cdots, k+1$,
corresponding to global minimums of the total energy.
As proven by Stieltjes, there is a unique configuration with
$-1<x_{1}<\cdots<x_{k}<1$ when $\gamma=0$ which corresponds to the
zeros of the Jacobi polynomial $P_{k}^{(\alpha-1,\beta-1)}(x)$. When $\gamma> 0$, however,
one can verify numerically as done in [15] that the range of the positions of these
positive unit charges tends to be along the entire half line except the two
singular points $\pm 1$  with
$\{x_{1},\cdots,x_{k}\}\in (-\infty,-1)~ \bigcup~(-1,+1)$.
Let the positions of these positive charges be arranged as
$x_{1}<x_{2}<\cdots<x_{k}$. The above restriction requires
that these positions must be within the intervals  $(-\infty,-1)$
and $(-1,+1)$ with
$-\infty<x_{1}<x_{2}<\cdots<x_{k_{1}}<-1<x_{k_{1}+1}<\cdots<x_{k}<+1$.
It follows from this that the total number of possible configurations is exactly
the number of ways to put the $k$ positive charges into the two intervals,
which is $\left(\matrix{k+m-1\cr k\cr}\right)=k+1$ for $m=2$.

Some explicit formulas for sums of powers of zeros of the new polynomials
can easily be derived. For example, summing Eq. (16) over $i$, we have

$$(\alpha-\beta)\sum_{i=1}^{k}{1\over{x_{i}^{2}-1}}+
(\alpha+\beta)\sum_{i=1}^{k}{x_{i}\over{x_{i}^{2}-1}}=-k\gamma,\eqno(25)$$
while multiplying Eq. (16) by $x_{i}$ and then summing over $i$, we get

$$(\alpha+\beta)\sum_{i=1}^{k}{1\over{x_{i}^{2}-1}}+
(\alpha-\beta)\sum_{i=1}^{k}{x_{i}\over{x_{i}^{2}-1}}+\gamma\sum_{i=1}^{k}x_{i}=-k(\alpha+\beta+k-1).
\eqno(26)$$
Eqs. (25) and (26) can be combined to express $\gamma\sum_{i=1}^{k}x_{i}$
related to the eigenenergy (14) as

$$\gamma\sum_{i=1}^{k}x_{i}={4\alpha\beta\over{\alpha+\beta}}\sum_{i=1}^{k}{1\over{1-x^{2}_{i}}}
-k(\alpha+\beta+k-1-\gamma{\alpha-\beta\over{\alpha+\beta}}).\eqno(27)$$

    Additional sum rules can be derived from the explicit form
of the new polynomial $y_{k}(x)$ by using the method shown in [37-39] for other
polynomials. Let the polynomial $y_{k}(x)$, up to a constant,
be expressed in terms of its zeros as

$$y_{k}(x)=\prod_{i=1}^{k}(x-x_{i}).\eqno(28)$$\\
Then, the expansion coefficients in (19) can be explicitly written as

$$b_{k}=1,~~b_{k-1}=-\sum_{i=1}^{k}x_{i},~~
b_{k-2}=\sum_{i<j}x_{i}x_{j},~~\cdots,~~
b_{0}=(-)^{k}\prod_{i=1}^{k}x_{i}.\eqno(29)$$\\
Directly using the four-term relation (21) with the explicit expressions shown
in (29), we have, for example,

$$f=\sum_{1\leq i<j\leq k}{2\over{x_{i}x_{j}}}+(\alpha-\beta-\gamma)\sum_{i=1}^{k}
{1\over{x_{i}}}=
\left(\sum_{i=1}^{k}{1\over{x_{i}}}\right)^{2}-(k-1)\sum_{i=1}^{k}{1\over{x_{i}^{2}}}
+(\alpha-\beta-\gamma)\sum_{i=1}^{k}
{1\over{x_{i}}}
\eqno(30)$$\\
for $k\geq2$, and
$$f=-\gamma\sum_{i=1}^{k}x_{i}-k(\alpha+\beta+k-1),\eqno(31)$$\\
which provides a relation among the eigenvalues $f$ in
the differential equation (18) and the eigenenergies (14) of the
special LMG model. Eq. (31) can easily be verified from the results
shown in Table 1.

Sum rules of other higher order
powers and inverse powers of zeros of the new polynomials
may also be derived from the four-term relation (21)
by using explicit expressions shown in (29). Eq. (31)
clearly shows that $f=-k(\alpha+\beta+k-1)$ corresponding
to the Jacobi polynomial with $y_{k}(x)=P^{(\alpha-1,\beta-1)}_{k}(x)$
when $\gamma=0$.

\section{Summary and discussions}

In this paper, new polynomials,
similar to but distinct from those
of the Heine-Stieltjes type, that are associated with a
special LMG model corresponding to the
standard two-site Bose-Hubbard model are derived
based on the Stieltjes correspondence.
It follows that the eigenvalues and corresponding
eigenstates of the special LMG model can be determined
from the zeros of these polynomials. Further, if these polynomials
can be derived recursively, it also follows that it should be much easier to
determine zeros from the polynomials, e. g. the case
considered in [40], than to solve a set
of BAEs with a set of variables because there is only one
variable in the polynomials.
This conclusion applies equally
well to other quantum many-body systems, such as
atomic-molecular Bose-Einstein condensates [41],
the heteronuclear molecular Bose-Einstein condensate model [42],
and the nuclear
pairing problem, when the corresponding polynomials can be
obtained.
It is well known that the Jacobi polynomials $P^{(\alpha,\beta)}_{k}(x)$
are orthogonal with respect to the measure $d\mu(x)=(x-1)^{\alpha}(x+1)^{\beta}dx$
on the interval $x\in (-1,+1)$. Although the orthogonality of the new polynomials
is not addressed, we conjuncture that, similar to the Lam\'{e}
polynomials analyzed in [43-44], there is no sequence of
the new polynomials with $\gamma\neq0$ orthogonal with respect to any measure
supported on the interval $(-\infty, 1)$ when $\gamma\neq 0$.
A rigorous proof of the latter, which requires further work,
is beyond the scope of the present study.
Furthermore, following Stieltjes early work, a one-dimensional
classical electrostatic analogue corresponding to the special LMG model
is established. This shows that any possible configuration of equilibrium
positions of the charges in the electrostatic problem corresponds
uniquely to one set of roots of the BAEs for the LMG model,
and that the number of possible configurations of equilibrium
positions of the charges equals exactly to the number of energy levels
in the LMG model.
Sum rules of powers and inverse powers of zeros of the new polynomials
related to  eigenenergies of the LMG model
are also considered. More complicated sum rules and their relations
may be obtained similarly. The results by extension clearly show a new link
between Bethe ansatz type solutions of large class of quantum many-body problems
and a class of polynomials satisfying second-order differential equations, and
thus opens a new way to get solutions of these BAEs from
zeros of these polynomials.

\vskip .5cm
One of the authors (P. F.) is grateful to Prof. Dr. Hans Volkmer
for providing information about his work on the
Heine-Stieltjes polynomials and related helpful discussions.
The authors are also grateful to the referees for providing
us relevant references to the subject and helpful comments.
Support from the U.S. National Science Foundation
(PHY-0500291 \& OCI-0904874), the Southeastern Universities Research Association, the
Natural Science Foundation of China (11175078),  the Doctoral Program Foundation
of State Education Ministry of China (20102136110002),
and the LSU--LNNU joint research program (9961) is
acknowledged.

\end{document}